\documentclass{article}
\usepackage{amssymb}

\usepackage{amsmath}
\usepackage{graphicx}
\usepackage{sw20jart}

\def\mail{maymin pmaymin@ltcm.com}
\def\prange{0 0}
\input ICCSsty.tex
\input{tcilatex}
\input tcilatex
\begin{document}

\title{Programming complex systems }
\author{Philip Maymin \\
%EndAName
Long-Term Capital Management}
\date{September 21, 1997}

\begin{abstract}
{Classical programming languages cannot model essential elements of complex
systems such as true random number generation. This paper develops a formal
programming language called the lambda-q calculus that addresses the
fundamental properties of complex systems. This formal language allows the
expression of quantumized algorithms, which are extensions of randomized
algorithms in that probabilities can be negative, and events can cancel out.
An illustration of the power of quantumized algorithms is the ability to
efficiently solve the satisfiability problem, something that many believe is
beyond the capability of classical computers. This paper proves that the
lambda-q calculus is not only capable of solving satisfiability but can also
simulate such complex systems as quantum computers. Since satisfiability is
believed to be beyond the capabilities of quantum computers, the lambda-q
calculus may be strictly stronger.}
\end{abstract}

\maketitle

\section{Introduction}

The purpose of this paper is to introduce a formalism for expressing models
of complex systems. The end result is that modelling any complex system such
as human society, evolution, or particle interactions, may be reduced to a
programming problem.

In addition to the modelling functionalities it provides, a programmable
complex system also allows us to see, in its specification, what the
distilled and essential elements of a complex system are. In particular, as
we will see, interactions like those in a cellular automaton need not be
explicit in the formalism, as they may be simulated.

Classical programming languages are not strong enough to model complex
systems. They do not allow for randomized events and are completely
predictable and deterministic, features rarely found in complex systems.
Some problems that may be quickly solved on quantum computers, which is a
complex system, have no known quick solutions on classical computers or with
classical programming languages.

In this paper we extend the $\lambda $-calculus, the logical foundation of
classical programming languages. The first extension, the $\lambda ^{p}$%
-calculus, is a new calculus introduced here for expressing randomized
functions. Randomized functions, instead of having a unique output for each
input, return a distribution of results from which we sample once. The $%
\lambda ^{p}$-calculus then provides a formal method for computing
distributions. More useful, however, would be the ability to compute
conditional distributions. The second extension, the $\lambda ^{q}$%
-calculus, is a new calculus introduced here for expressing quantumized
functions. Quantumized functions also return a distribution of results,
called a \emph{superposition}, from which we sample once, but $\lambda ^{q}$%
-terms have signs, and identical terms with opposite signs are removed
before sampling from the result. Quantumized functions can then compute
conditional distributions. The effect is that of applying some filter to a
superposition to adjust each of the probabilities according to its fitness.
One example is the quick solution of satisfiability:\ by merely filtering
out the logical mappings of variables that do not satisfy the given formula,
we are left only with satisfying mappings, if any. The $\lambda ^{q}$%
-calculus is the most general of the three calculi.

One of the results of this paper is that the $\lambda ^{q}$-calculus is at
least as powerful as quantum computers. Although much research has been done
on the hardware of quantum computation (c.f. \cite{deutsch 85}, \cite
{deutsch 89}, \cite{simon}), none has focused on formalizing the software.
Quantum Turing machines \cite{deutsch 85} have been introduced but there has
been no quantum analogue to Church's $\lambda $-calculus. The $\lambda $%
-calculus has served as the basis for many programming languages since it
was introduced by Alonzo Church \cite{church} in 1936. It and other
classical calculi make the implicit assumption that a term may be
innocuously observed at any point. Such an assumption is hard to separate
from a system of rewriting rules because to rewrite a term, you must have
read it. One of the goals of these calculi is to make observation explicit.

The $\lambda ^{p}$- and the $\lambda ^{q}$-calculi allow the expression of
algorithms that exist and operate in the Heisenberg world of \emph{potentia} 
\cite{heisenberg} but whose results are observed. To this end, collections
(distributions and superpositions) should be thought of with the following
intuition. A collection is a bunch of terms that co-exist in the same place
but are not aware of each other. Thus, a collection of three terms takes up
no more space than a collection of two terms. A physical analogy is the
ability of a particle to be in a superposition of states. When the
collection is observed, at most one term in each collection will be the
result of the observation. The key point is that in neither calculus can one
write a term that can determine if it is part of a collection, how big the
collection is, or even if its argument is part of a collection. Despite this
inability, the $\lambda ^{q}$-calculus is powerful enough to efficiently
solve problems such as satisfiability that are typically believed to be
beyond the scope of classical computers.

\section{The Lambda Calculus}

This section is a review of the $\lambda $-calculus and a reference for
later calculi. For more details see e.g. \cite{barendregt}.

The $\lambda $-calculus is a calculus of functions. Any computable
single-argument function can be expressed in the $\lambda $-calculus. Any
computable multiple-argument function can be expressed in terms of
computable single-argument functions. The $\lambda $-calculus is useful for
encoding functions of arbitrary arity that return at most one output for
each input. In particular, the $\lambda $-calculus can be used to express
any (computable) \emph{algorithm}. The definition of algorithm is usually
taken to be Turing-computable.

\subsection{Syntax}

The following grammar specifies the syntax of the $\lambda $-calculus. 
\begin{equation}
\begin{tabular}{|ll|}
\hline
$
\begin{array}{ll}
x & \in \text{\emph{Variable}} \\ 
M & \in \text{\emph{LambdaTerm}} \\ 
w & \in \text{\emph{Wff}}
\end{array}
$ & $
\begin{array}{l}
\text{Variables} \\ 
\text{Terms} \\ 
\text{Well-formed formulas}
\end{array}
$ \\ 
&  \\ 
$
\begin{array}{lll}
M & ::= & x \\ 
& \,\,\,| & M_{1}M_{2} \\ 
& \,\,\,| & \lambda x.M \\ 
&  &  \\ 
w & ::= & M_{1}=M_{2}
\end{array}
$ & $
\begin{array}{l}
\text{variable} \\ 
\text{application} \\ 
\text{abstraction} \\ 
\\ 
\text{well-formed formula}
\end{array}
$ \\ \hline
\end{tabular}
\label{lambda syntax}
\end{equation}

To be strict, the subscripts above should be removed (e.g., the rule for
well-formed formulas should read\ $w::=M=M$) because $M_{1}$ and $M_{2}$ are
not defined. However, we will maintain this incorrect notation to emphasize
that the terms need not be identical.

With this abuse of notation, we can easily read the preceding definition as:
a $\lambda $-term is a variable, or an application of two terms, or the
abstraction of a term by a variable. A well-formed formula of the $\lambda $%
-calculus is a $\lambda $-term followed by the equality sign followed by a
second $\lambda $-term.

We also adopt some syntactic conventions. Most importantly, parentheses
group subexpressions. Application is taken to be left associative so that
the term $MNP$ is correctly parenthesized as $\left( MN\right) P$ and not as 
$M\left( NP\right) .$ The scope of an abstraction extends as far to the
right as possible, for example up to a closing parenthesis, so that the term 
$\lambda x.xx$ is correctly parenthesized as $\left( \lambda x.xx\right) $
and not as $\left( \lambda x.x\right) x.$

\subsection{Substitution}

We will want to substitute arbitrary $\lambda $-terms for variables. We
define the substitution operator, notated$~M\left[ N/x\right] $ and read ``$%
M $ with all free occurences of $x$ replaced by $N$.'' The definition of the
free and bound variables of a term are standard. The set of free variables
of a term $M$ is written $FV\left( M\right) $. There are six rules of
substitution, which we write for reference. 
\begin{equation}
\begin{array}{l}
1.\;x\left[ N/x\right] \equiv N \\ 
2.\;y\left[ N/x\right] \equiv y\,\,\text{for variables }y\not{\equiv}x \\ 
3.\;\left( PQ\right) \left[ N/x\right] \equiv \left( P\left[ N/x\right]
\right) \left( Q\left[ N/x\right] \right) \\ 
4.\;\left( \lambda x.P\right) \left[ N/x\right] \equiv \lambda x.P \\ 
5.\;\left( \lambda y.P\right) \left[ N/x\right] \equiv \lambda y.\left(
P\left[ N/x\right] \right) \,\,\text{if } 
\begin{array}{l}
y\not{\equiv}x\text{ and} \\ 
y\notin FV\left( N\right)
\end{array}
\\ 
6.\;\left( \lambda y.P\right) \left[ N/x\right] \equiv \lambda z.\left(
P\left[ z/y\right] \left[ N/x\right] \right) \\ 
\,\,\,\,\,\,\,\,\,\,\,\,\,\,\,\,\,\,\,\,\,\,\text{if } 
\begin{array}{l}
y\not{\equiv}x\text{,} \\ 
y\in FV\left( N\right) \text{, and} \\ 
z\notin FV(P)\bigcup FV\left( N\right)
\end{array}
\end{array}
\label{substitution}
\end{equation}

This definition will be extended in both subsequent calculi.

\subsection{Reduction}

\label{notions of reduction}The concept of \emph{reduction} seeks to
formalize rewriting rules. Given a relation $R$ between terms, we may define
the one-step reduction relation, notated$~\rightarrow _{R},$ that is the
contextual closure of $R.$ We may also define the reflexive, transitive
closure of the one-step reduction relation, which we call $R$-reduction and
notate$~\twoheadrightarrow _{R},$ and the symmetric closure of $R$%
-reduction, called $R$-interconvertibility and notated$~=_{R}.$

The essential notion of reduction for the $\lambda $-calculus is called $%
\beta $-reduction. It is based on the $\beta $-relation, which is the
formalization of function invocation. 
\begin{equation}
\beta \triangleq \left\{ 
\begin{array}{c}
\left( \left( \lambda x.M\right) N,M\left[ N/x\right] \right) \,\,\, \\ 
\text{s.t.}\,\,M,N\in LambdaTerm,\,x\in Variable
\end{array}
\right\}  \label{beta}
\end{equation}

There is also the $\alpha $-relation that holds of terms that are identical
up to a consistent renaming of variables. 
\begin{equation}
\alpha \triangleq \left\{ 
\begin{array}{c}
\left( \lambda x.M,\lambda y.M\left[ y/x\right] \right) \,\,\, \\ 
\text{s.t.}\,\,\,M\in LambdaTerm,\,y\notin FV\left( M\right)
\end{array}
\right\}
\end{equation}
We will use this only sparingly.

\subsection{Evaluation Semantics}

By imposing an evaluation order on the reduction system, we are providing
meaning to the $\lambda $-terms. The evaluation order of a reduction system
is sometimes called an operational semantics or an evaluation semantics for
the calculus. The evaluation relation is typically denoted $\rightsquigarrow
.$

We use call-by-value evaluation semantics. A \emph{value }is the result
produced by the evaluation semantics. Call-by-value semantics means that the
body of an abstraction is not reduced but arguments are evaluated before
being passed into abstractions.

There are two rules for the call-by-value evaluation semantics of the $%
\lambda $-calculus. 
\begin{eqnarray*}
&&\frac {}{v\rightsquigarrow v}\text{(Refl)\qquad \qquad (for }v\text{ a
value)} \\
&&\frac{M\rightsquigarrow \lambda x.P\quad N\rightsquigarrow N^{\prime
}\quad P\left[ N^{\prime }/x\right] \rightsquigarrow v}{MN\rightsquigarrow v}%
\text{(Eval)}
\end{eqnarray*}

\subsection{Reference Terms}

The following $\lambda $-terms are standard and are provided as reference
for later examples.

Numbers are represented as Church numerals. 
\begin{eqnarray}
\underline{0} &\equiv &\lambda x.\lambda y.y \\
\underline{n} &\equiv &\lambda x.\lambda y.x^{n}y
\end{eqnarray}

\noindent where the notation $x^{n}y$ means $n$ right-associative
applications of $x$ onto $y.$ It is abbreviatory for the term $
\begin{array}{l}
\underbrace{x(x(\cdots (x}y))) \\ 
\,n\text{ times}
\end{array}
.$ When necessary, we can extend Church numerals to represent both positive
and negative numbers. For the remainder of the terms, we will not provide
definitions. The predecessor of Church numerals is written $\underline{\text{%
P}}.$ The successor is written $\underline{\text{S}}.$

The conditional is written $\underline{\text{IF}}.$ If its first argument is
truth, written $\underline{\text{T}},$ then it returns its second argument.
If its first argument is falsity, written $\underline{\text{F}},$ then it
returns its third argument. A typical predicate is $\underline{\text{0?}}$
which returns $\underline{\text{T}}$ if its argument is the Church numeral $%
\underline{\text{0}}$ and $\underline{\text{F}}$ if it is some other Church
numeral.

The fixed-point combinator is written $\underline{\text{Y}}.$ The primitive
recursive function-building term is written $\underline{\text{PRIM-REC}}$
and it works as follows. If the value of a function $f$ at input $n$ can be
expressed in terms of $n-1$ and $f\left( n-1\right) ,$ then that function $f$
is primitive recursive, and it can be generated by providing $\underline{%
\text{PRIM-REC}}$ with the function that takes the inputs $n-1$ and $f\left(
n-1\right) $ to produce $f\left( n\right) $ and with the value of $f$ at
input $0.$ For example, the predecessor function for Church numerals can be
represented as $\underline{\text{P}}\equiv \underline{\text{PRIM-REC}}%
\,\left( \lambda x.\lambda y.x\right) \,\underline{\text{0}}.$

\section{The Lambda-P Calculus}

The $\lambda ^{p}$-calculus is an extension of the $\lambda $-calculus that
permits the expression of \emph{randomized }algorithms. In contrast with a
computable algorithm which returns at most one output for each input, a
randomized algorithm returns a \emph{distribution} of answers from which we
sample. There are several advantages to randomized algorithms.

\begin{enumerate}
\item  Randomized algorithms can provide truly random number generators
instead of relying on pseudo-random number generators that work only because
the underlying pattern is difficult to determine.

\item  Because they can appear to generate random numbers arbitrarily,
randomized algorithms can model random processes.

\item  Given a problem of finding a suitable solution from a set of
possibilities, a randomized algorithm can exhibit the effect of choosing
random elements and testing them. Such algorithms can sometimes have an 
\emph{expected }running time which is considerably shorter than the running
time of the computable algorithm that tries every possibility until it finds
a solution.
\end{enumerate}

\subsection{Syntax}

\label{section:lambda-p syntax}The following grammar describes the $\lambda
^{p}$-calculus. 
\begin{equation}
\begin{tabular}{|ll|}
\hline
$
\begin{array}{ll}
x & \in \text{\emph{Variable}} \\ 
M & \in \text{\emph{LambdaPTerm}} \\ 
w & \in \text{\emph{WffP}}
\end{array}
$ & $
\begin{array}{l}
\text{Variables} \\ 
\text{Terms} \\ 
\text{Well-formed formulas}
\end{array}
$ \\ 
&  \\ 
$
\begin{array}{lll}
M & ::= & x \\ 
& \,\,\,| & M_{1}M_{2} \\ 
& \,\,\,| & \lambda x.M \\ 
& \,\,\,| & M_{1},M_{2} \\ 
&  &  \\ 
w & ::= & M_{1}=M_{2}
\end{array}
$ & $
\begin{array}{l}
\text{variable} \\ 
\text{application} \\ 
\text{abstraction} \\ 
\text{collection} \\ 
\\ 
\text{well-formed formula}
\end{array}
$ \\ \hline
\end{tabular}
\newline
\label{lambda-p syntax}
\end{equation}

Since this grammar differs from the $\lambda $-calculus only in the addition
of the fourth rule for terms, all $\lambda $-terms can be viewed as $\lambda
^{p}$-terms. A $\lambda ^{p}$-term may be a collection of a term and another
collection, so that a $\lambda ^{p}$-term may actually have many nested
collections.

We adhere to the same parenthesization and precedence rules as the $\lambda $%
-calculus with the following addition:\ collection is of lowest precedence
and the comma is right associative. This means that the expression $\lambda
x.x,z,y$ is correctly parenthesized as $\left( \lambda x.x\right) ,(z,y)$.

We introduce abbreviatory notation for collections. Let us write $\left[
M_{i}^{i\in S}\right] $ for the collection of terms $M_{i}$ for all $i$ in
the finite, ordered set $S$ of natural numbers. We will write $a..b$ for the
ordered set $\left( a,a+1,\ldots ,b\right) .$ In particular, $\left[
M_{i}^{i\in 1..n}\right] $ represents $M_{1},M_{2},\ldots ,M_{n}$ and $%
\left[ M_{i}^{i\in n..1}\right] $ represents $M_{n},M_{n-1},\ldots ,M_{1}$.
More generally, let us allow multiple iterators in arbitrary contexts. Then,
for instance, 
\[
\left[ \lambda x.M_{i}^{i\in 1..n}\right] \equiv \lambda x.M_{1},\lambda
x.M_{2},\ldots ,\lambda x.M_{n} 
\]
and 
\[
\left[ M_{i}^{i\in 1..m}N_{j}^{j\in 1..n}\right] \equiv 
\begin{array}{c}
M_{1}N_{1},M_{1}N_{2},\ldots ,M_{1}N_{n}, \\ 
M_{2}N_{1},M_{2}N_{2},\ldots ,M_{2}N_{n}, \\ 
\vdots \\ 
M_{m}N_{1},M_{m}N_{2},\ldots ,M_{m}N_{n}
\end{array}
. 
\]

Note that $\left[ \lambda x.M_{i}^{i\in 1..n}\right] $ and $\lambda x.\left[
M_{i}^{i\in 1..n}\right] $ are not the same term. The former is a collection
of abstractions while the latter is an abstraction with a collection in its
body. Finally, we allow this notation to hold of non-collection terms as
well by identifying $\left[ M_{i}^{i\in 1..1}\right] $ with $M_{1}$ even if $%
M_{1}$ is not a collection. To avoid confusion, it is important to
understand that although this ``collection'' notation can be used for
non-collections, we do not extend the definition of the word \emph{%
collection. }A \emph{collection} is still the syntactic structure defined in
grammar (\ref{lambda-p syntax}).

With these additions, every term can be written in this bracket form. In
particular, we can write a collection as $\left[ \left[ M_{i}^{i\in
S_{i}}\right] _{j}^{j\in S}\right] ,$ or a collection of collections.
Unfortunately, collections can be written in a variety of ways with this
notation. The term $M,N,P$ can be written as $\left[ M_{i}^{i\in
1..3}\right] $ if $M_{1}\equiv M$ and $M_{2}\equiv N$ and $M_{3}\equiv P;$
as $\left[ M_{i}^{i\in 1..2}\right] $ if $M_{1}\equiv M$ and $M_{2}\equiv
N,P;$ or as $\left[ M_{i}^{i\in 1..1}\right] $ if $M_{1}\equiv M,N,P.$
However, it cannot be written as $\left[ M_{i}^{i\in 1..4}\right] $ for any
identification of the $M_{i}.$ This observation inspires the following
definition.

\begin{definition}
\label{dfn: cardinality}The \emph{cardinality} of a term $M,$ notated$%
~\left| M\right| ,$ is that number $k$ for which $\left[ M_{i}^{i\in
1..k}\right] \equiv M$ for some identification of the $M_{i}$ but $\left[
M_{i}^{i\in 1..\left( k+1\right) }\right] \not{\equiv}M$ for any
identification of the $M_{i}$.
\end{definition}

\noindent Note that the cardinality of a term is always strictly positive.

\subsection{Syntactic Identities}

We define substitution of terms in the $\lambda ^{p}$-calculus as an
extension of substitution of terms in the $\lambda $-calculus. In addition
to the substitution rules of the $\lambda $-calculus, we introduce one for
collections. 
\begin{equation}
\left( P,Q\right) \left[ N/x\right] \equiv \left( P\left[ N/x\right]
,Q\left[ N/x\right] \right)  \label{substitution-p}
\end{equation}

We identify terms that are collections but with a possibly different
ordering. We also identify nested collections with the top-level collection.
The motivation for this is the conception that a collection is an unordered
set of terms. Therefore we will not draw a distinction between a set of
terms and a set of a set of terms.

We adopt the following axiomatic judgement rules. 
\begin{eqnarray*}
&&\dfrac {}{M,N\equiv N,M}\text{(ClnOrd)} \\
&&\dfrac {}{\left( M,N\right) ,P\equiv M,(N,P)}\text{(ClnNest)}
\end{eqnarray*}

With these axioms, ordering and nesting become innocuous. As an example here
is the proof that $A,(B,C),D\equiv A,C,B,D.$ For clarity, we parenthesize
fully and underline the affected term in each step. 
\[
\begin{array}{llll}
\underline{A,((B,C),D)} & \equiv & ((\underline{B,C}),D),A & \text{(ClnOrd)}
\\ 
& \equiv & (\underline{(C,B),D}),A & \text{(ClnOrd)} \\ 
& \equiv & \underline{(C,(B,D)),A} & \text{(ClnNest)} \\ 
& \equiv & A,(C,(B,D)) & \text{(ClnOrd)}
\end{array}
\]

It can be shown that ordering and parenthesization are irrelevant in
general. Aside, it no longer matters that we took the comma to be right
associative since any arbitrary parenthesization of a collection does not
change its syntactic structure.

Because of this theorem, we can alter the abbreviatory notation and allow
arbitrary unordered sets in the exponent. This allows us to write, for
instance, $\left[ M_{i}^{i\in 1..n-\{j\}}\right] \equiv M_{1},M_{2},\ldots
,M_{j-1},M_{j+1},\ldots ,M_{n}$ where $a..b$ is henceforth taken to be the
unordered set $\left\{ a,a+1,\ldots ,b\right\} $ and the subtraction in the
exponent represents set difference.

This also subtly alters the definition of \emph{cardinality }(\ref{dfn:
cardinality}). Whereas before the cardinality of a term like $\left(
x,y\right) ,z$ was 2, because of this theorem, it is now 3.

We may now also introduce a further abbreviation. We let $\left[ \left(
M_{i}:n_{i}\right) \right] $ be a rewriting of the term $\left[ N_{i}^{i\in
I}\right] $ such each of the $M_{i}$ are distinct and the integer $n_{i}$
represents the count of each $M_{i}$ in $\left[ N_{i}^{i\in I}\right] .$

\subsection{Reductions}

The relation of collection application is called the $\gamma $-relation. It
holds of a term that is an application at least one of whose operator or
operand is a collection, and the term that is the collection of all possible
pairs of applications.{}

\begin{equation}
\gamma ^{p}\triangleq \left\{ 
\begin{array}{l}
\left( \left[ M_{i}^{i\in 1..m}\right] \left[ N_{j}^{j\in 1..n}\right]
,\left[ M_{i}^{i\in 1..m}N_{j}^{j\in 1..n}\right] \right) \\ 
\text{s.t. }M_{i},N_{j}\in LambdaPTerm,\,m>1\text{ or }n>1
\end{array}
\right\}  \label{gamma-p}
\end{equation}

\noindent We will omit the superscript except to disambiguate from the $%
\gamma $-relation of the $\lambda ^{q}$-calculus.

It can be shown that the $\gamma $-relation is Church-Rosser and that all
terms have $\gamma $-normal forms. Therefore, we may write $\gamma \left(
M\right) $ for the $\gamma $-normal form of $M.$

We extend the $\beta $-relation to apply to collections. 
\begin{equation}
\beta ^{p}\triangleq \left\{ 
\begin{array}{l}
\left( \left( \lambda x.M\right) \left[ N_{i}^{i\in S}\right] ,\left[
M\left[ N_{i}^{i\in S}/x\right] \right] \right) \\ 
\text{s.t. }M\text{, }\left[ N_{i}^{i\in S}\right] \in LambdaPTerm,\,x\in
Variable
\end{array}
\right\}  \label{beta-p}
\end{equation}
where $\left[ M\left[ N_{i}^{i\in S}/x\right] \right] $ is the collection of
terms $M$ with $N_{i}$ substituted for free occurrences of $x$ in $M,$ for $%
i\in S.$

\subsection{Evaluation Semantics}

We extend the call-by-value evaluation semantics of the $\lambda $-calculus.
We modify the definition of a value $v$ to enforce that $v$ has no $\gamma $%
-redexes.

\begin{eqnarray*}
&&\dfrac {}{v\rightsquigarrow v}\text{(Refl)\qquad \qquad (for }v\text{ a
value)} \\
&&\dfrac{\gamma \left( M\right) \rightsquigarrow \lambda x.P\quad \gamma
\left( N\right) \rightsquigarrow N^{\prime }\quad \gamma \left( P\left[
N^{\prime }/x\right] \right) \rightsquigarrow v}{MN\rightsquigarrow v}\text{%
(Eval)} \\
&&\dfrac{\gamma \left( M\right) \rightsquigarrow v_{1}\quad \gamma \left(
N\right) \rightsquigarrow v_{2}}{\left( M,N\right) \rightsquigarrow \left(
v_{1},v_{2}\right) }\text{(Coll)}
\end{eqnarray*}

\subsection{Observation}

\label{lambda-p observation}We define an observation function $\Theta $ from 
$\lambda ^{p}$-terms to $\lambda $-terms. We employ the random number
generator $RAND$, which samples one number from a given set of numbers. 
\begin{eqnarray}
\Theta \left( x\right) &=&x \\
\Theta \left( \lambda x.M\right) &=&\lambda x.\Theta \left( M\right) \\
\Theta \left( M_{1}M_{2}\right) &=&\Theta \left( M_{1}\right) \Theta \left(
M_{2}\right) \\
\Theta \left( M\equiv \left[ M_{i}^{i\in 1..\left| M\right| }\right] \right)
&=&M_{RAND(1..\left| M\right| )}
\end{eqnarray}

\noindent The function $\Theta $ is total because every $\lambda ^{p}$-term
is mapped to a $\lambda $-term. Note that for an arbitrary term $T$ we may
write $\Theta \left( T\right) =T_{RAND(S)}$ for some possibly singleton set
of natural numbers $S$ and some collection of terms $\left[ T_{i}^{i\in
S}\right] .$

We can show that observing a $\lambda ^{p}$-term is statistically
indistinguishable from observing its $\gamma $-normal form.

\subsection{Observational Semantics}

We provide another type of semantics for the $\lambda ^{p}$-calculus called
its \emph{observational semantics.} A formalism's observational semantics
expresses the computation as a whole:\ preparing the input, waiting for the
evaluation, and observing the result. The observational semantics relation
between $\lambda ^{p}$-terms and $\lambda $-terms is denoted$~\multimap $.
It is given by a single rule for the $\lambda ^{p}$-calculus. 
\begin{equation}
\frac{M\rightsquigarrow v\quad \Theta \left( v\right) =N}{M\multimap N}\text{%
(ObsP)}  \label{obs-p}
\end{equation}

\subsection{Examples}

A useful term of the $\lambda ^{p}$-calculus is a random number generator.
We would like to define a term that takes as input a numeral \underline{$n$}
and computes a collection of numerals from \underline{$0$} to \underline{$n$}%
. This can be represented by the following primitive recursive $\lambda ^{p}$%
-term. 
\begin{equation}
\underline{\text{R}}\equiv \underline{\text{PRIM-REC}}\,\left( \lambda
k.\lambda p.\left( k,p\right) \right) \,\underline{\text{0}}
\end{equation}

\noindent Then for instance $\underline{\text{R}}\,\underline{\text{3}}%
=\left( \underline{3},\underline{2},\underline{1},\underline{0}\right) .$

The following term represents a random walk. Imagine a man that at each
moment can either walk forward one step or backwards one step. If he starts
at the point $0$, after $n$ steps, what is the distribution of his position? 
\begin{equation}
\underline{\text{W}}\equiv \underline{\text{PRIM-REC}}\,\left( \lambda
k.\lambda p.\left( \underline{\text{P}}p,\underline{\text{S}}p\right)
\right) \,\underline{\text{0}}
\end{equation}

\noindent We assume we have extended Church numerals to negative numbers as
well. This can be easily done by encoding it is a pair. We will show some of
the highlights of the evaluation of $\underline{\text{W}}\,\underline{\text{3%
}}.$ Note that $\underline{\text{W}}\,\underline{\text{1}}=\left( \underline{%
-1},\underline{1}\right) .$%
\begin{equation}
\begin{array}{lll}
\underline{\text{W}}\,\underline{\text{3}} & = & \underline{\text{P}}\left( 
\underline{\text{W}}\,\underline{\text{2}}\right) ,\underline{\text{S}}%
\left( \underline{\text{W}}\,\underline{\text{2}}\right) \\ 
& = & \underline{\text{P}}\left( \underline{\text{P}}\left( \underline{\text{%
W}}\,\underline{\text{1}}\right) ,\underline{\text{S}}\left( \underline{%
\text{W}}\,\underline{\text{1}}\right) \right) ,\underline{\text{S}}\left( 
\underline{\text{P}}\left( \underline{\text{W}}\,\underline{\text{1}}\right)
,\underline{\text{S}}\left( \underline{\text{W}}\,\underline{\text{1}}%
\right) \right) \\ 
& = & \underline{\text{P}}\left( \underline{\text{P}}\left( \underline{-1},%
\underline{1}\right) ,\underline{\text{S}}\left( \underline{-1},\underline{1}%
\right) \right) ,\underline{\text{S}}\left( \underline{\text{P}}\left( 
\underline{-1},\underline{1}\right) ,\underline{\text{S}}\left( \underline{-1%
},\underline{1}\right) \right) \\ 
& = & \underline{\text{P}}\left( \left( \underline{-2},\underline{0}\right)
,\left( \underline{0},\underline{2}\right) \right) ,\underline{\text{S}}%
\left( \left( \underline{-2},\underline{0}\right) ,\left( \underline{0},%
\underline{2}\right) \right) \\ 
& = & \left( \left( \underline{-3},\underline{-1}\right) ,\left( \underline{%
-1},\underline{1}\right) \right) ,\left( \left( \underline{-1},\underline{1}%
\right) ,\left( \underline{1},\underline{3}\right) \right) \\ 
& \equiv & \left( \underline{-3},\underline{-1},\underline{-1},\underline{1},%
\underline{-1},\underline{1},\underline{1},\underline{3}\right)
\end{array}
\end{equation}

\noindent Observing $\underline{\text{W}}\,\underline{\text{3}}$ yields $%
\underline{-1}$ with probability $\frac{3}{8},$ $\underline{1}$ with
probability $\frac{3}{8},$ $\underline{-3}$ with probability $\frac{1}{8}$,
and $\underline{3}$ with probability $\frac{1}{8}.$

\section{The Lambda-Q Calculus}

The $\lambda ^{q}$-calculus is an extension of the $\lambda ^{p}$-calculus
that allows easy expression of \emph{quantumized }algorithms. A quantumized
algorithm differs from a randomized algorithm in allowing negative
probabilities and in the way we sample from the resulting distribution.

Variables and abstractions in the $\lambda ^{q}$-calculus have \emph{phase}.
The phase is nothing more than a plus or minus sign, but since the result of
a quantumized algorithm is a distribution of terms with phase, we call such
a distribution by the special name \emph{superposition}. The major
difference between a superposition and a distribution is the observation
procedure. Before randomly picking an element, a superposition is
transformed into a distribution by the following two-step process. First,
all terms in the superposition that are identical except with opposite phase
are cancelled. They are both simply removed from the superposition. Second,
the phases are stripped to produce a distribution. Then, an element is
chosen from the distribution randomly, as in the $\lambda ^{p}$-calculus.

The words \emph{phase} and \emph{superposition} come from quantum physics.
An electron is in a superposition if it can be in multiple possible states.
Although the phases of the quantum states may be any angle from $0{{}^{\circ
}}$ to $360{{}^{\circ }}$, we only consider binary phases. Because we use
solely binary phases, we will use the words \emph{sign} and \emph{phase }%
interchangeably in the sequel.

A major disadvantage of the $\lambda ^{p}$-calculus is that it is impossible
to compress a collection. Every reduction step at best keeps the collection
the same size. Quantumized algorithms expressed in the $\lambda ^{q}$%
-calculus, on the other hand, can do this as easily as randomized algorithms
can generate random numbers. That is, $\lambda ^{q}$-terms can contain
subterms with opposite signs which will be removed during the observation
process.

\subsection{Syntax}

The following grammar describes the $\lambda ^{q}$-calculus. 
\begin{equation}
\begin{tabular}{|ll|}
\hline
$
\begin{array}{ll}
S & \in \text{\emph{Sign}} \\ 
x & \in \text{\emph{Variable}} \\ 
M & \in \text{\emph{LambdaQTerm}} \\ 
w & \in \text{\emph{WffQ}}
\end{array}
$ & $
\begin{array}{l}
\text{Sign, or phase} \\ 
\text{Variables} \\ 
\text{Terms} \\ 
\text{Well-formed formulas}
\end{array}
$ \\ 
&  \\ 
$
\begin{array}{lll}
S & ::= & + \\ 
& \,\,\,| & - \\ 
&  &  \\ 
M & ::= & Sx \\ 
& \,\,\,| & M_{1}M_{2} \\ 
& \,\,\,| & S\lambda x.M \\ 
& \,\,\,| & M_{1},M_{2} \\ 
&  &  \\ 
w & ::= & M_{1}=M_{2}
\end{array}
$ & $
\begin{array}{l}
\text{positive} \\ 
\text{negative} \\ 
\\ 
\text{signed variable} \\ 
\text{application} \\ 
\text{signed abstraction} \\ 
\text{collection} \\ 
\\ 
\text{well-formed formula}
\end{array}
$ \\ \hline
\end{tabular}
\newline
\label{lambda-q syntax}
\end{equation}

Terms of the $\lambda ^{q}$-calculus differ from terms of the $\lambda ^{p}$%
-calculus only in that variables and abstractions are \emph{signed}, that
is, they are preceded by either a plus (+)\ or a minus (-)\ sign. Just as $%
\lambda $-terms could be read as $\lambda ^{p}$-terms, we would like $%
\lambda ^{p}$-terms to be readable as $\lambda ^{q}$-terms. However, $%
\lambda ^{p}$-terms are unsigned and cannot be recognized by this grammar.

Therefore, as is traditionally done with integers, we will omit the positive
sign. An unsigned term in the $\lambda ^{q}$-calculus is abbreviatory for
the same term with a positive sign. With this convention, $\lambda ^{p}$%
-terms can be seen as $\lambda ^{q}$-terms all of whose signs are positive.
Also, so as not to confuse a negative sign with subtraction, we will write
it with a logical negation sign\ ($\lnot $). With these two conventions, the 
$\lambda ^{q}$-term $+\lambda x.+x-\!x$ is written simply $\lambda x.x\lnot
x.$

Finally, we adhere to the same parenthesization and precedence rules as the $%
\lambda ^{p}$-calculus. In particular, we continue the use of the
abbreviatory notations $\left[ M_{i}^{i\in S}\right] $ and $\left[ \left(
M_{i}:n_{i}\right) \right] $ for collections of terms. In addition, we can
also $\left[ \left( M_{i}:n_{i}\right) \right] $ as $\left[ \left(
M_{i}:a_{i},b_{i},n_{i}\right) \right] $ such that $M_{i}\not{\equiv}M_{j}$
and $M_{i}\not{\equiv}\overline{M_{j}}$ for $i\neq j,$ all of the $M_{i}$
are of positive sign, the integer $a_{i}$ denotes the count of $M_{i},$ the
integer $b_{i}$ denotes the count of $\overline{M_{i}},$ and $%
n_{i}=a_{i}-b_{i}.$

\subsection{Syntactic Identities}

We will call two terms \emph{opposites} if they differ only in sign.

We define substitution of terms in the $\lambda ^{q}$-calculus as a
modification of substitution of terms in the $\lambda ^{p}$-calculus. We
rewrite the seven rules of the $\lambda ^{p}$-calculus to take account of
the signs of the terms. First, we introduce the function notated by sign
concatenation, defined by the following rule in our abbreviatory
conventions. 
\begin{equation}
\lnot \lnot \mapsto \epsilon
\end{equation}

\noindent We also note that the concatenation of a sign $S$ with $\epsilon $
is just $S$ again. Now we can use this function in the following
substitution rules. 
\begin{equation}
\begin{array}{l}
1.\;\left( Sx\right) \left[ N/x\right] \equiv SN \\ 
2.\;\left( Sy\right) \left[ N/x\right] \equiv Sy\,\,\text{for variables }y%
\not{\equiv}x \\ 
3.\;\left( PQ\right) \left[ N/x\right] \equiv \left( P\left[ N/x\right]
\right) \left( Q\left[ N/x\right] \right)  \\ 
4.\;\left( S\lambda x.P\right) \left[ N/x\right] \equiv S\lambda x.P \\ 
5.\;\left( S\lambda y.P\right) \left[ N/x\right] \equiv S\lambda y.\left(
P\left[ N/x\right] \right) \,\,\text{if }
\begin{array}{l}
y\not{\equiv}x\text{, and} \\ 
y\notin FV\left( N\right) 
\end{array}
\\ 
6.\;\left( S\lambda y.P\right) \left[ N/x\right] \equiv S\lambda z.\left(
P\left[ z/y\right] \left[ N/x\right] \right)  \\ 
\,\,\,\,\,\,\,\,\,\,\,\,\,\,\,\,\,\,\,\,\,\text{if }
\begin{array}{l}
y\not{\equiv}x\text{,} \\ 
y\in FV\left( N\right) \text{, and} \\ 
z\notin FV(P)\bigcup FV\left( N\right) 
\end{array}
\\ 
7.\;\left( P,Q\right) \left[ N/x\right] \equiv \left( P\left[ N/x\right]
,Q\left[ N/x\right] \right) 
\end{array}
\label{substitution-q}
\end{equation}

The use of the sign concatenation function is hidden in rule (1). Consider $%
\left( \lnot x\right) \left[ \lnot \lambda y.y/x\right] \equiv \lnot \lnot
\lambda y.y.$ This is not a $\lambda ^{q}$-term by grammar (\ref{lambda-q
syntax}) but applying the sign concatenation function yields the term $%
\lambda y.y$.

\subsection{Reduction}

The $\gamma $-relation of the $\lambda ^{q}$-calculus is of the same form as
that of the $\lambda ^{p}$-calculus.

\begin{equation}
\gamma ^{q}\triangleq \left\{ 
\begin{array}{l}
\left( \left[ M_{i}^{i\in 1..m}\right] \left[ N_{j}^{j\in 1..n}\right]
,\left[ M_{i}^{i\in 1..m}N_{j}^{j\in 1..n}\right] \right) \\ 
\text{s.t. }M_{i},N_{j}\in LambdaQTerm,\,m>1\text{ or }n>1
\end{array}
\right\}  \label{gamma-q}
\end{equation}
We omit the superscript when it is clear if the terms under consideration
are $\lambda ^{p}$-terms or $\lambda ^{q}$-terms. We still write $\gamma
\left( M\right) $ for the $\gamma $-normal form of $M.$

We extend the $\beta $-relation to deal properly with signs.

\begin{equation}
\beta ^{q}\triangleq \left\{ 
\begin{array}{l}
\left( \left( S\lambda x.M\right) N,SM\left[ N/x\right] \right) \\ 
\text{s.t. }S\in \text{\emph{Sign}},\text{and }S\lambda x.M,N\in LambdaQTerm
\end{array}
\right\}
\end{equation}

\subsection{Evaluation Semantics}

We modify the call-by-value evaluation semantics of the $\lambda ^{p}$%
-calculus.

\noindent 
\begin{eqnarray*}
&&\frac {}{v\rightsquigarrow v}\text{(Refl)\qquad \qquad (for }v\text{ a
value)} \\
&&\frac{\gamma \left( M\right) \rightsquigarrow S\lambda x.P\,\,\,\,\gamma
\left( N\right) \rightsquigarrow N^{\prime }\,\,\,\,\,\gamma \left( SP\left[
N^{\prime }/x\right] \right) \rightsquigarrow v}{MN\rightsquigarrow v}\text{%
(Eval)} \\
&&\frac{\gamma \left( M\right) \rightsquigarrow v_{1}\quad \gamma \left(
N\right) \rightsquigarrow v_{2}}{\left( M,N\right) \rightsquigarrow \left(
v_{1},v_{2}\right) }\text{(Coll)}
\end{eqnarray*}

\subsection{Observation}

\label{lambda-q observation}We define an observation function $\Xi $ from $%
\lambda ^{q}$-terms to $\lambda $-terms as the composition of a function $%
\Delta $ from $\lambda ^{q}$-terms to $\lambda ^{p}$-terms with the
observation function $\Theta $ from $\lambda ^{p}$-terms to $\lambda $-terms
defined in (\ref{lambda-p observation}). Thus, $\Xi =\Theta \circ \Delta $
where we define $\Delta $ as follows. 
\begin{eqnarray}
\Delta \left( Sx\right) &=&x \\
\Delta \left( S\lambda x.M\right) &=&\lambda x.\Delta \left( M\right) \\
\Delta \left( M_{1}M_{2}\right) &=&\Delta \left( M_{1}\right) \Delta \left(
M_{2}\right) \\
\Delta \left( \left[ M_{i}:a_{i},b_{i},n_{i}\right] \right) &=&\left[ \Delta
\left( M_{i}^{i\in \left\{ i\,\,\,|\,\,\,n_{i}\neq 0\right\} }:\left|
n_{i}\right| \right) \right]  \label{Collection case for delta}
\end{eqnarray}

Note that unlike the observation function $\Theta $ of the $\lambda ^{p}$%
-calculus, the observation function $\Xi $ of the $\lambda ^{q}$-calculus is
not total. For example, $\Xi \left( x,\lnot x\right) $ does not yield a $%
\lambda $-term because $\Delta \left( x,\lnot x\right) $ is the empty
collection, which is not a $\lambda ^{p}$-term.

Although observing a $\lambda ^{p}$-term is statistically indistinguishable
from observing its $\gamma $-normal form, observing a $\lambda ^{q}$-term
is, in general, statistically distinguishable from observing its $\gamma $%
-normal form.

\subsection{Observational Semantics}

The observational semantics for the $\lambda ^{q}$-calculus is similar to
that of the $\lambda ^{p}$-calculus (\ref{obs-p}). It is given by a single
rule. 
\begin{equation}
\frac{M\rightsquigarrow v\quad \Xi \left( v\right) =N}{M\multimap N}\text{%
(ObsQ)}  \label{obs-q}
\end{equation}

\subsection{Examples}

We provide one example. We show how satisfiability may be solved in the $%
\lambda ^{q}$-calculus. We assume possible solutions are encoded some way in
the $\lambda ^{q}$-calculus and there is a term $\underline{\text{CHECK}_{f}}
$ that checks if the fixed Boolean formula $f$ is satisfied by a particular
truth assignment, given as the argument. The output from this is a
collection of $\underline{\text{T}}$ (truth) and $\underline{\text{F}}$
(falsity) terms. We now present a term that will effectively remove all of
the $\underline{\text{F}}$ terms. It is an instance of a more general
method. 
\begin{equation}
\underline{\text{REMOVE-F}}\equiv \lambda x.\,\underline{\text{IF}}%
\,x\,x\,\left( x,\lnot x\right)
\end{equation}

We give an example evaluation. 
\begin{equation}
\begin{array}{lll}
\underline{\text{REMOVE-F}}\,\left( \underline{\text{F}},\underline{\text{T}}%
,\underline{\text{F}}\right) & \equiv & \left( \lambda x.\,\underline{\text{%
IF}}\,x\,x\,\left( x,\lnot x\right) \right) \left( \underline{\text{F}},%
\underline{\text{T}},\underline{\text{F}}\right) \\ 
& \rightarrow _{\gamma } & \left( 
\begin{array}{l}
\left( \lambda x.\,\underline{\text{IF}}\,x\,x\,\left( x,\lnot x\right)
\right) \underline{\text{F}}, \\ 
\left( \lambda x.\,\underline{\text{IF}}\,x\,x\,\left( x,\lnot x\right)
\right) \underline{\text{T}}, \\ 
\left( \lambda x.\,\underline{\text{IF}}\,x\,x\,\left( x,\lnot x\right)
\right) \underline{\text{F}}
\end{array}
\right) \\ 
& \twoheadrightarrow _{\beta } & \left( \left( \underline{\text{F}},\lnot 
\underline{\text{F}}\right) ,\underline{\text{T}},\left( \underline{\text{F}}%
,\lnot \underline{\text{F}}\right) \right) \\ 
& \equiv & \left( \underline{\text{F}},\lnot \underline{\text{F}},\underline{%
\text{T}},\underline{\text{F}},\lnot \underline{\text{F}}\right)
\end{array}
\end{equation}

\noindent Observing the final term will always yield $\underline{\text{T}}.$
Note that the drawback to this method is that if $f$ is unsatisfiable then
the term will be unobservable. Therefore, when we insert a distinguished
term into the collection to make it observable, we risk observing that term
instead of $\underline{\text{T}}.$ At worst, however, we would have a
fifty-fifty chance of error.

Specifically, consider what happens when the argument to $\underline{\text{%
REMOVE-F}}$ is a collection of $\underline{\text{F}}^{\prime }$s$.$ Then $%
\underline{\text{REMOVE-F}}\,\underline{\text{F}}=\left( \underline{\text{F}}%
,\lnot \underline{\text{F}}\right) .$We insert $\underline{\text{I}}\equiv
\lambda x.x$ which, if we observe, we take to mean that either $f$ is
unsatisfiable or we have bad luck. Thus, we observe the term $\left( 
\underline{\text{I}},\underline{\text{F}},\lnot \underline{\text{F}}\right)
. $ This will always yield $\underline{\text{I}}.$ However, we cannot
conclude that $f$ is unsatisfiable because, in the worst case, the term may
have been $\left( \underline{\text{I}},\underline{\text{REMOVE-F}}\,%
\underline{\text{T}}\right) =\left( \underline{\text{I}},\underline{\text{T}}%
\right) $ and we may have observed $\underline{\text{I}}$ even though $f$
was satisfiable. We may recalculate until we are certain to an arbitrary
significance that $f$ is not satisfiable.

Therefore, applying $\underline{\text{REMOVE-F}}$ to the results of $%
\underline{\text{CHECK}_{f}}$ and then observing the result will yield $%
\underline{\text{T}}$ only if $f$ is satisfiable.

\section{Simulation to quantum computers}

We show that the $\lambda ^{q}$-calculus can efficiently simulate the \emph{%
one-dimensional partitioned quantum cellular automata} (1d-PQCA) defined in 
\cite{Watrous 1995}. By the equivalence of 1d-PQCA and quantum Turing
machines\ (QTM) proved in \cite{Watrous 1995}, the $\lambda ^{q}$-calculus
can efficiently simulate QTM.

To show that 1d-PQCA can be efficiently simulated by the $\lambda ^{q}$%
-calculus, we need to exhibit a $\lambda ^{q}$-term $M$ for a given 1d-PQCA $%
A$ such that $A$ after $k$ steps is in the same superposition as $M$ after $%
P\left( k\right) $ steps, with $P$ a polynomial.

We assume for now that the 1d-PQCA has transition amplitudes not over the
complex numbers, but over the positive and negative rationals. It has been
shown \cite{bernstein/vazirani} that this is equivalent to the general model
in QTM.

To express $A$ in $M$, we need to do the following things.

\begin{enumerate}
\item  Translate states of $A$ into $\lambda ^{q}$-terms that can be
compared\ (e.g. into Church numerals).

\item  Translate the acceptance states and the integer denoting the
acceptance cell into $\lambda ^{q}$-terms.

\item  Create a $\lambda ^{q}$-term $\mathbf{P}$ to mimic the operation of
the permutation $\sigma .$

\item  Translate the local transition function into a transition term. For
1d-PQCA this means translating the matrix $\Lambda $ into a term $\mathbf{L}$
comparing the initial state with each of the possible states and returning
the appropriate superposition.

\item  Determine an injective mapping of configurations of $A$ and
configurations of $M$.
\end{enumerate}

Although we will not write down $M$ in full, we note that within $M$ are the
mechanisms described above that take a single configuration, apply $\mathbf{P%
}$, and return the superposition as described by $\mathbf{L}.$

We recall that the contextual closure of the $\beta ^{q}$-relation is such
that $M,N\rightarrow _{\beta }M^{\prime },N^{\prime }$ where $M\rightarrow
_{\beta }M^{\prime }$and $N\rightarrow _{\beta }N^{\prime }.$ Thus there is
parallel reduction within superpositions. By inspection of the mechanisms
above it follows that $k$ steps of $A$ is equivalent to a polynomial of $k$
steps of $M$.

Steps 1, 2, and 3 are straightforward. Then for step 5, the $\lambda ^{q}$%
-superposition $\left[ \left( M_{i}:a_{i},b_{i},n_{i}\right) \right] $ (let $%
n=\sum n_{i}$) will be equivalent to the 1d-PQCA-superposition $\sum \frac{%
n_{i}}{n}\left| c\left( M_{i}\right) \right\rangle ,$ where $c$ takes $%
\lambda ^{q}$-terms and translates them into 1d-PQCA configurations.
Essentially this means stripping off everything other than the data, that is
to say, the structure containing the contents. Note that $c$ is not itself a 
$\lambda ^{q}$-term. It merely performs a fixed syntactic operation,
removing extraneous information such as $\mathbf{P}$ and $\mathbf{L,}$ and
translating the Church numerals that represent states into the 1d-PQCA
states. This is injective because the mapping from states of $A$ into
numerals is injective. Thus, step 5 is complete.

Step 4 requires translating the $\Lambda $ matrix into a matrix of whole
numbers, and translating an arbitrary 1d--PQCA superposition into a $\lambda
^{q}$-superposition. The latter is done merely by multiplying each of the
amplitudes by the product of the denominators of all of the amplitudes, to
get integers. We call the product of the denominators here $d$. We perform a
similar act on the $\Lambda $ matrix, multiplying each element by the
product of all of the denominators of $\Lambda .$ We call this constant $b.$
Then we have that $T=b\Lambda $ is a matrix over integers. This matrix can
be considered notation for the $\lambda ^{q}$-term that checks if a given
state is a particular state and returns the appropriate superposition. For
instance, if 
\[
\Lambda =\left( 
\begin{array}{ll}
\frac{2}{3} & \frac{1}{3} \\ 
0 & 1
\end{array}
\right) 
\]
then 
\[
T=b\Lambda =9\Lambda =\left( 
\begin{array}{ll}
6 & 3 \\ 
0 & 9
\end{array}
\right) 
\]
which we can consider as alternate notation for 
\begin{eqnarray*}
\mathbf{Q} &\equiv &\lambda s.\text{\textbf{\ IF }(\textbf{EQUAL }}s\text{%
\textbf{1}) (\textbf{1,1,1,1,1,1,2,2,2)}} \\
&&\text{(\textbf{IF} (\textbf{EQUAL\ }}s\text{\textbf{2}) (\textbf{%
2,2,2,2,2,2,2,2,2}))}
\end{eqnarray*}
Then it follows that if $c$ is a superposition of configuration of $A$,
applying $\Lambda $ $k$ times results in the same superposition as applying $%
T$ $k$ times to the representation of $c$ in the $\lambda ^{q}$-calculus.

\section{Conclusion}

We have seen two new formalisms. The $\lambda ^{p}$-calculus allows
expression of randomized algorithms. The $\lambda ^{q}$-calculus allows
expression of quantumized algorithms. In these calculi, observation is made
explicit, and the notion of superposition common to quantum physics is
formalized for algorithms.

This work represents a new direction of research. Just as the $\lambda $%
-calculus found many uses in classical programming languages, the $\lambda
^{p}$-calculus and the $\lambda ^{q}$-calculus may help discussion of
randomized and quantum programming languages.

It should not be difficult to see that the $\lambda ^{p}$-calculus can
simulate a probabilistic Turing machine and we have shown that the $\lambda
^{q}$-calculus can simulate a quantum Turing machine\ (QTM). However, as we
have shown, the $\lambda ^{q}$-calculus can efficiently solve NP-complete
problems such as satisfiability, while there is widespread belief (e.g. \cite
{bennett}) that QTM\ cannot efficiently solve satisfiability. Thus, the
greater the doubt that QTM cannot solve NP-complete problems, the greater
the justification in believing that the $\lambda ^{q}$-calculus is strictly
stronger than QTM.

It should also follow that a probabilistic Turing machine can
(inefficiently) simulate the $\lambda ^{p}$-calculus. However, it is not
obvious that a quantum Turing machine can simulate the $\lambda ^{q}$%
-calculus. An answer to this question will be interesting. If quantum
computers can simulate the $\lambda ^{q}$-calculus efficiently, then the $%
\lambda ^{q}$-calculus can be used as a programming language directly. As a
byproduct, satisfiability will be efficiently and physically solvable. If
quantum computers cannot simulate the $\lambda ^{q}$-calculus efficiently,
knowing what the barrier is may allow the formulation of another type of
computer that can simulate it.

\section{Acknowledgements}

Thanks to Stuart Shieber for helpful comments.

\end{document}